\documentclass[10pt, conference, compsocconf]{IEEEtran}

\ifCLASSINFOpdf
\else
\fi

\usepackage{amsmath,amssymb,amsfonts, mathtools}
\usepackage{algorithm, algpseudocode}
\usepackage{graphicx}
\usepackage{textcomp}
\usepackage{xcolor}
\usepackage{booktabs} 
\usepackage{hyperref}
\usepackage{amsmath}
\usepackage{relsize}
\usepackage{setspace}
\usepackage{subfig}

\usepackage{url}

\begin{document}
%
\title{Efficient Bitmap-based Indexing and Retrieval of Similarity Search Image Queries}


\author{\IEEEauthorblockN{Omid Jafari}
	\IEEEauthorblockA{\textit{Computer Science Department} \\
		\textit{New Mexico State University}\\
		Las Cruces, USA \\
		ojafari@nmsu.edu}
	\and
	\IEEEauthorblockN{Parth Nagarkar}
	\IEEEauthorblockA{\textit{Computer Science Department} \\
		\textit{New Mexico State University}\\
		Las Cruces, USA \\
		nagarkar@nmsu.edu}
	\and
	\IEEEauthorblockN{Jonathan Monta\~no}
	\IEEEauthorblockA{\textit{Department of Mathematical Sciences} \\
		\textit{New Mexico State University}\\
		Las Cruces, USA \\
		jmon@nmsu.edu}
}


%


\maketitle

\begin{abstract}
Finding similar images is a necessary operation in many multimedia applications. Images are often represented and stored as a set of high-dimensional features, which are extracted using localized feature extraction algorithms. 
Locality Sensitive Hashing is one of the most popular approximate processing techniques for finding similar points in high-dimensional spaces. Locality Sensitive Hashing (LSH) and its variants are designed to find similar points, but they are not designed to find objects (such as images, which are made up of a collection of points) efficiently. In this paper, we propose an index structure, Bitmap-Image LSH (\textit{bImageLSH}), for efficient processing of high-dimensional images. Using a real dataset, we experimentally show the performance benefit of our novel design while keeping the accuracy of the image results high.

\end{abstract}


\IEEEpeerreviewmaketitle

\section{Introduction}

Many large multimedia applications require efficient processing of nearest neighbor (or similarity search) queries. Often, multimedia, such as images, are represented and stored as a collection of high-dimensional features, which are extracted using feature-extraction algorithms such as SIFT \cite{Sift:Lowe2004}, SURF \cite{Surf}, etc. Traditional tree-based index structures
suffer from the popular \textit{curse of dimensionality}. One solution to address the \textit{curse of dimensionality} is to search for approximate results instead of exact results. 
One of the most popular solutions for approximate query processing in high-dimensional spaces is \textit{Locality Sensitive Hashing} \cite{Gionis:1999:SSH:645925.671516}. 

\subsection{Locality Sensitive Hashing}

Locality Sensitive Hashing (LSH) maps high-dimensional data to lower dimensional representations by using \textit{random} hash functions. The intuition behind LSH is that points that are nearby in the original high-dimensional data are mapped to the same (or neighboring) hash buckets in the hash functions (lower-dimensional projected space) with a high probability, and points that are far apart in the original space are mapped to the same hash buckets in the hash functions with a low probability. Many variants of LSH that focus on improving the search performance and/or the search accuracy have been proposed (please see the survey paper \cite{Hashing:Survey} for more details). 

\subsubsection{Motivation for using LSH-based Techniques}

There are two main benefits of using 
LSH for finding similar points in high-dimensional spaces: sub-linear query performances (in terms of the data size) and proven theoretical guarantees on the accuracy of the returned results. 
Since LSH uses data-independent random hash functions (i.e. data characteristics such as data distribution are not needed to create these hash functions), the creation of these hash functions is a simple process that takes negligible time. In applications where the data is changing (i.e. the data distribution is changing) and/or newer data is incoming at a fast pace, the time taken to create or update hash functions during runtime can significantly impact the query performance. Random hash functions, as opposed to \textit{learned} hash functions, do not require any modification during runtime. While the original LSH index structure suffered from large index sizes (in order to obtain a high query accuracy) \cite{Lv:2007:MLE:1325851.1325958}, state-of-the-art LSH techniques \cite{Gan:2012:LHS:2213836.2213898, Huang:2015:QLH:2850469.2850470} have alleviated this issue by using advanced methods such as \textit{Collision Counting} and \textit{Virtual Rehashing}. Thus, owing to their small index sizes, and most importantly, lack of any required expensive online processes, in this paper, we propose an efficient index structure, Bitmap-Image LSH (\textit{bImageLSH}), that is built upon existing LSH techniques for faster similarity search queries on images.

\subsection{Motivation of our work: Drawbacks of LSH-based techniques for finding Similar Images}
\label{sec:motv}
Images are often represented and stored by a collection of high-dimensional \textit{descriptors} (which consists of multiple \textit{features}). E.g. a SIFT \cite{Sift:Lowe2004} descriptor consists of 128 features, and an image consists of multiple SIFT descriptors (In the Wang dataset (Section \ref{sec:exp}), every image is represented by an average of 695 128-dimensional descriptors). Different images can be represented by a different number of descriptors. 
If an image $X_1$ has more details than image $X_2$, then SIFT will extract more descriptors for $X_1$ than it will for $X_2$. 

If a user wants to find similar images to a given query image, nearest-neighbor queries (or top-$k$ queries) have to be performed for every individual descriptor representing the query image. Once the results of the individual descriptor queries are found, popular voting count techniques such as the Borda Count method \cite{borda:doi:10.1177/0192512102023004002} or its variants \cite{borda:PEREZ2011951} are used to aggregate results of the descriptor queries to find similar images \cite{borda:7536065, Arora:2018:HPS:3204028.3228393}. An overall score is assigned to each image object in the database based on the depth of its descriptors in the top-$k'$ results of the descriptor queries (we refer the reader to \cite{Arora:2018:HPS:3204028.3228393} for a formal definition of this score). Once these scores are calculated for every image in the database, the top-$k$ images are returned. Here, $k$ is the user-specified number of desired similar images and $k'$ is a constant chosen to find closest descriptors for the individual descriptor queries (\cite{Arora:2018:HPS:3204028.3228393} chooses $k'$ as 100). Higher the $k'$, the accuracy of the final $k$ results will be higher but the performance will be slower and vice-versa. 

There are three main drawbacks of this approach: 1) all descriptor queries of the image query are executed
independently of each other, 2) there is no theoretical guarantee
for the accuracy of the returned top-$k$ result images, and 3) there are no performance optimizations for the overall image query execution since the descriptor queries are executed independently. Given an image query, if a descriptor query takes
too long to execute as compared to others, then the overall processing time is negatively affected. This is especially
disadvantageous if the top-$k$ images could have been found before all descriptor queries have finished execution.

\subsection{Contributions of our work}
In this paper, we introduce an index structure, \textit{bImageLSH}, that can efficiently processing similarity search queries for images. In the context of \textit{Locality Sensitive Hashing}, our work is the first attempt at efficiently finding similar images from a real image dataset. We present two stopping conditions that aid our novel design. We present a novel bitmap-based optimization for pruning images that will not be returned in the result set, thus saving IO costs. We will leverage this framework to eventually prove theoretical bounds on our stopping conditions to provide a robust and an efficient index structure for finding similar images. 

\section{Related Work}

The goal of our proposed index structure, Bitmap-Image LSH (\textit{bImageLSH}), is to efficiently find similar images to a given image query (which has been converted into a set of high-dimensional descriptor queries). Existing LSH-based techniques (except PSLSH \cite{Nagarkar:CIKM} and QWLSH \cite{Jafari:2019:QCI:3323873.3325048}) \cite{Gionis:1999:SSH:645925.671516, Lv:2007:MLE:1325851.1325958, Gan:2012:LHS:2213836.2213898, Huang:2015:QLH:2850469.2850470} focus on optimizing a single descriptor query (instead of a set of descriptor queries). 
The problem formulation of \cite{Nagarkar:CIKM} is different because it focuses on returning points that satisfy a certain user-defined percentage of the descriptor queries from a query workload.
\cite{Jafari:2019:QCI:3323873.3325048} focuses on building a model based on the data characteristics to determine how to optimally utilize the cache during query processing. The main drawback of these approaches is that they require prior information in terms of models from existing datasets. \textit{bImageLSH} does not require any prior knowledge about the data in order to efficiently find similar images.
In \cite{yu2012cb}, the authors propose to improve the performance of deletion of traditional LSH algorithms by using compressed bitmaps, which is different from our bitmap-based optimization.

There have been several works that have defined voting-based similarity/distance measures between two images 
\cite{Jegou:2010:IBL:1718320.1718326, Jegou:2009}. 
\textit{bImageLSH} is orthogonal to these approaches. 
Our eventual goal (Section \ref{sec:future}) is to return results with a theoretical guarantee efficiently.

\section{Problem Specification}

In this section, we formally describe the problem we solve in this paper. Given a multidimensional database $\mathcal{D}$, $\mathcal{D}$ consists of $n$ $d$-dimensional points that belong to a bounded multidimensional space $\mathcal{R}^d$. Each $d$-dimensional point $x_i$ is associated with an image object $X_j$ s.t. multiple points are \textit{associated} with a single image object. There are $S$ image objects in the database ($1 \leq S \leq n$), and for each image object $X_j$, $desc(X_j)$ denotes the set of points (descriptors) that are associated with $X_j$, and $|desc(X_j)|$ denotes the number of points that are associated with $X_j$. 

In this paper, our goal is to solve the $k$-NN version of the $c$-approximate nearest neighbor problem \cite{Gan:2012:LHS:2213836.2213898}. Let us denote the distance between two image objects $X_1$ and $X_2$ by $dist(X_1, X_2)$. Let $sim(X_1, X_2) = \frac{|\{x_1 \in desc(X_1), x_2 \in desc(X_2) : \; ||x_1, x_2|| \leq R\}|}{|desc(X_1)|.|desc(X_2)|}$ and $sim(X_1, X_2) = 1 - dist(X_1, X_2)$. $R$ is the given radius and $||x_1, x_2||$ denotes the Euclidean distance between two points $x_1$ and $x_2$. For a given query image $Q$, an image $X_j$ is a $c$-approximate nearest neighbor of query $Q$ if the distance between $Q$ and $X_j$ is at most $c$ times the distance between $Q$ and its true (or exact) nearest neighbor, $X_j^*$, i.e. $dist(Q, X_j) \leq c\times dist(Q, X_j^*)$, where $c>1$ is an \textit{approximation ratio}. Similarly, the $k$-NN version of this problem states that we want to find $k$ images that are respectively the $c$-approximate nearest images of the exact $k$-NN images of $Q$. 

\section{Design of bitmap-Image LSH (bImageLSH)}

The main goal of \textit{bImageLSH} is to efficiently return top-k images for a given query image without affecting the accuracy of the result. During query processing, instead of executing the query descriptors of query image $Q$ independently, we execute them one at a time in each projection (hash function). Given a query image $Q$, in the context of Locality Sensitive Hashing, we define a score called \textit{Collision Index} ($ci$) for each image that determines how close two images are based on the number of points between the two images that are considered as candidates (i.e. the collision counts between the points of the two images was greater than the collision threshold $l$ \cite{Gan:2012:LHS:2213836.2213898}): $ci(Q, X_j) = \frac{|\{q \in desc(Q), x_i \in desc(X_j) : \; cc(q, x_i) \geq l\}|}{|desc(Q)|.|desc(X_j)|}$. Similarly, we define an approximate distance between two images called $cDist(Q, X_j)$ which is equal to $1- \frac{|\{q \in desc(Q), x_i \in desc(X_j) : \; ||q, x_i|| \leq c.R\}|}{|desc(Q)|.|desc(X_j)|}$.

The Collision Index between two images depends on how many nearby points are considered as candidates between
the two images. Thus, in turn, the accuracy of the collision index depends on the accuracy of the collision counting process (i.e. if two points are nearby, then the collision count between these two points should be greater than the collision
threshold $l$). The values for the Collision Indexes between the query image and the other images in the database are dependent on the diversity of the database. If the database consists of very similar images, then more number of points belonging to different images will be similar (i.e. considered as candidates) to the query image, and hence the collision index values will be higher.
Hence we define a constant threshold $\Gamma$ such that if the collision index of two images $(ci(Q, X_j))$ is greater than or equal to $\Gamma$, we consider the two images as $\Gamma$-close Images (i.e. $X_j$ is considered as a candidate for query image $Q$). These two scores help us define two stopping conditions: \textit{S1}) At any radius, there exists at least $k$ images that are $\Gamma$-close to query image $Q$ and whose $cDist(Q, X_j) \geq \Gamma$, \textit{and S2}) At any radius, for each query descriptor $q$ in $Q$, $k' + v'$ candidate points are found (where $v'$ is the number of false positive points allowed \cite{Gan:2012:LHS:2213836.2213898}). If $\Gamma$ is estimated to be too high, then \textit{bImageLSH} will stop at the same radius as the naive borda count process (Section \ref{sec:motv}). 

One of the dominant costs in query processing is accessing the index files from the secondary storage. In the process of \textit{Virtual Rehashing}, the LSH algorithm increases the radius exponentially every time if sufficient results are not found. With each increase in radius, a new set of index files needs to be accessed to see which points collide with the point query. We propose a new strategy to reduce this cost: We first classify the images in the database into 3 categories: \textit{Useful} images, \textit{Maybe-useful} images, and \textit{Useless} images. At the beginning, all images are assigned to the \textit{Maybe-useful} category since we do not know which images will be close to the query image. The index files contain points (which belong to different images). We assign an upper bound score (which is the $\Gamma$-threshold) and a lower bound score (which is equal to the \textit{Collision Index} of the $k+v$th image, where $v$ is the allowed number of false positive images). Thus, at a particular radius, if the \textit{Collision Index} of an image is greater than 0 but lower than the lower bound score, then we can safely ignore this image for any further processing (by classifying it as a \textit{Useless} image). If an image $X_j$ is $\Gamma$-close to the image query and $cDist(Q, X_j) \geq \Gamma$, then we classify it as a \textit{Useful} image. 

The challenge here is that the index files contain points, and not images. One solution is to additionally represent each index file as a compressed bitmap, where the length of the bitmap is equal to the number of images in the database. If a point in the index file belongs to an image $X_j$ ($1 \leq j \leq S$), then the $j$th bit in the bitmap is turned to 1. This offline process can be done during index construction. During query processing, we store a single \textit{image bitmap} whose bits are turned to 1 if the corresponding image is classified as a \textit{Maybe-useful} image. For images that only belong to this category, further processing of points is necessary. In order to efficiently check whether an index file only contains points belonging to either of the other 2 categories (\textit{Useful} or \textit{Useless} images), we do a bitwise AND operation between the \textit{image bitmap} and the bitmap representing the index file. If the resultant bitmap consists of all 0 bits, then we know that the index file only contains points that belong to \textit{Useful} or \textit{Useless} image. Hence we do not need to bring the index file into the memory for further processing. We noticed that these index files always contain a small percentage of \textit{Useful} or \textit{Useless} images. Hence, we define a threshold, $UThres$, for minimum number of \textit{Maybe-useful} images that can exist in an index file. If the number of \textit{Maybe-useful} images in the index file is below $UThres$, then we ignore that index file. The benefit of using bitmaps is two-fold: 1) since index files contain only few elements, the bitmaps will be smaller since the compression ratio will be higher, and 2) doing a bitwise AND operation to find the resultant bitmap is a very fast operation. This small overhead of the AND operation is more beneficial than having to do an IO operation to bring the index file into the main memory from the secondary storage for processing.

\section{Preliminary Results}
\label{sec:exp}

We use a real image data set for our evaluation: \textbf{WangImage}\cite{wangdataset} This dataset consists of 695,672 128-dimensional SIFT descriptors belonging to 1000 images. These images belong to 10 different categories; each category has 100 similar images. All experiments were run on machines: Intel Core i7-6700, 16GB RAM, 2TB HDD, and Ubuntu 16.04. We used the state-of-the-art QALSH \cite{Huang:2015:QLH:2850469.2850470} as our base implementation. \textit{bImageLSH} can be implemented over any state-of-the-art LSH variant. QALSH stores all index files in memory. We modified the code such that the index files are stored on the secondary storage, and they are accessed from the secondary storage when needed. 
All codes were written in C++-11 (gcc v5.4 with the -O3 flag). We set $\Gamma = 0.0475\%$ and $UThres = 3\%$ in our experiments. 
Since there is no work that directly aims at solving our problem in the LSH domain, we compare our work with the following alternatives: 
\textbf{QALSH-Borda:} The top-100 results of the point queries are found using QALSH. The borda count process is applied to find the most similar images.

We evaluate the performance and accuracy of \textit{bImageLSH} using the following criteria:
\textbf{IndexIO: }The main dominant cost in LSH-based techniques is the index IO time. We observed that the index IO times were not consistent (i.e. running the same query, which needed the same amount of index IO, would return drastically different results, mainly because of disk cache and instruction cache issues). Hence, we report the savings in MB instead of the time taken to read these files. \textbf{Accuracy: }We run a top-20 image query on 5 randomly chosen images. For each image category in \textbf{WangImage} dataset, we have the ground truth of 100 images that belong to the category. If a returned result belongs to the category, we consider that result as a true positive. We define accuracy ($acc$) as: $acc = \frac{\# (true\_positives)}{\#(desired\_results)}$. By using compressed bitmaps, we add a storage overhead of $28.7\%$ on top of the existing QALSH index. 

\begin{figure}[!t]
	\centering
	\includegraphics[width=2.5in]{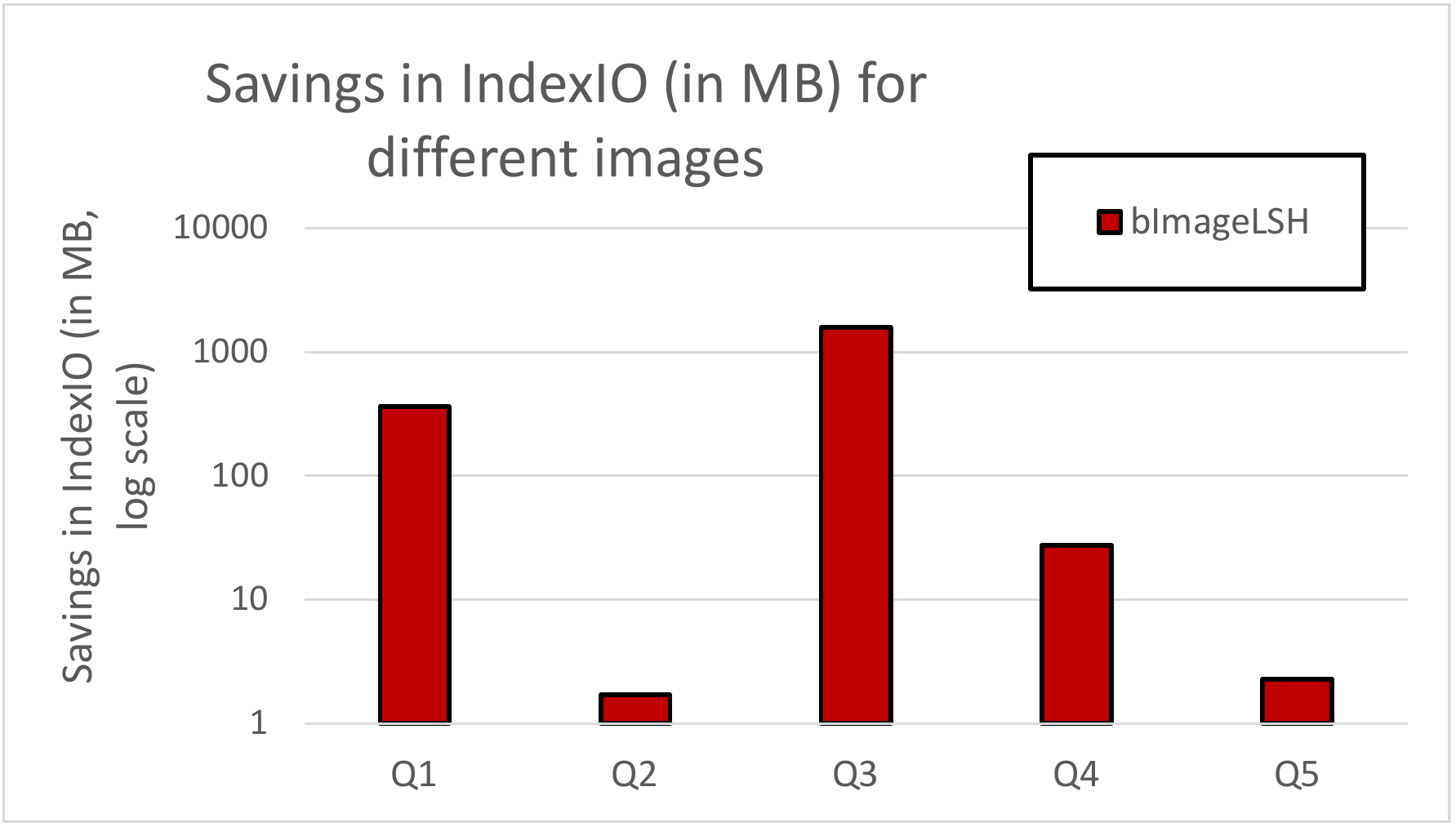}
	\caption{Savings in IndexIO (in MB)}
	\label{fig:IO}
\end{figure}

\begin{figure}[!t]
	\centering
	\includegraphics[width=2.5in]{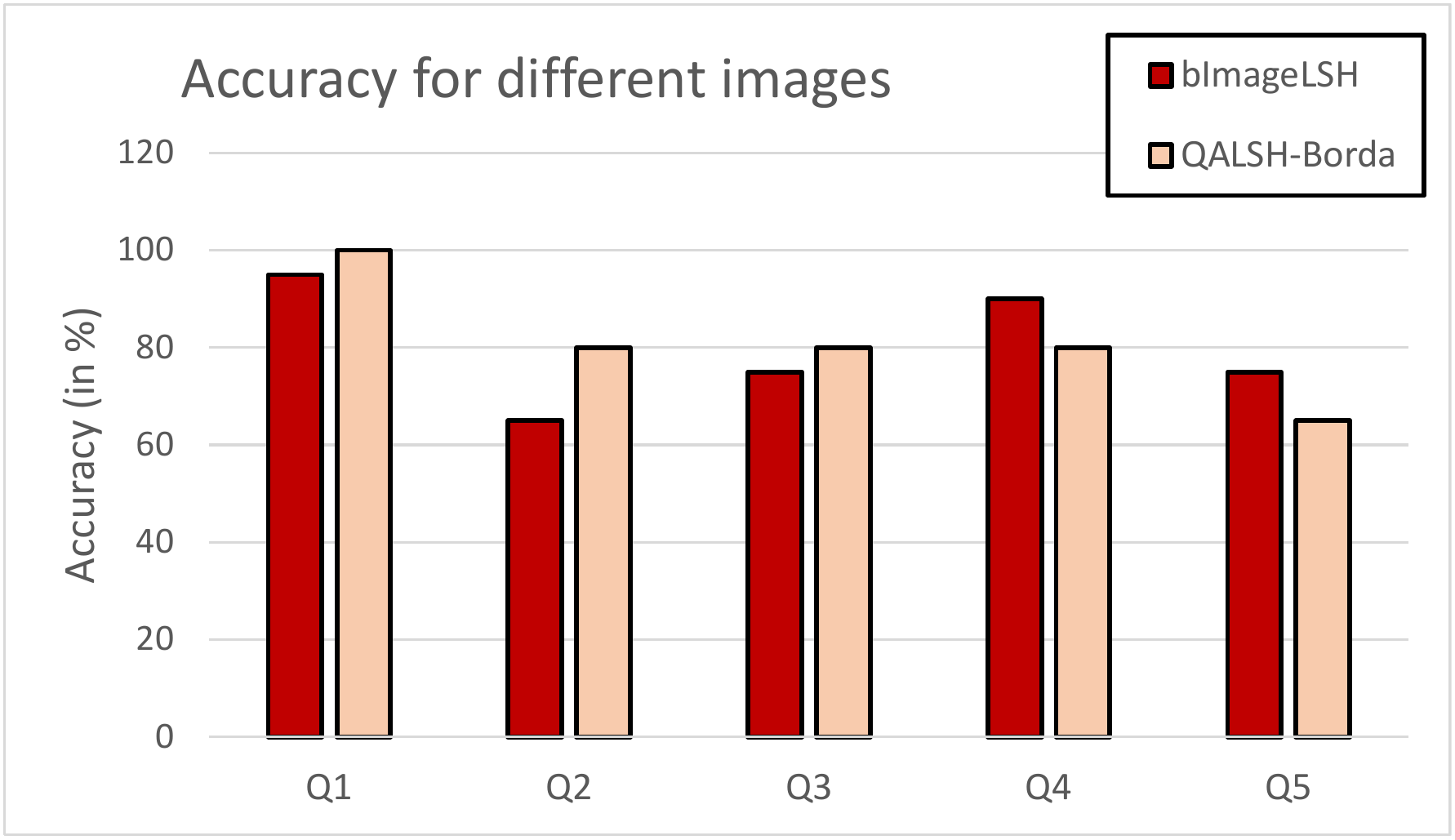}
	\caption{Comparison of Accuracy of the results}
	\label{fig:accuracy}
\end{figure}

From Figure \ref{fig:IO}, we can see that we save an average of 394.12 MB of IndexIO (per query image) due to our stopping conditions and bitmap-based pruning optimization. Figure \ref{fig:accuracy} shows that the accuracy of our results is not significantly affected. The $UThres$ threshold determines the trade-off between the amount of pruning and the accuracy of the results. In these experiments, we set $UThres$ to a conservative $3\%$ to keep the accuracy high. In future work, we will explore the impact of different values of $UThres$ on the accuracy of the results. 


\section{Future Work}
\label{sec:future}
In this paper, we introduced \textit{bImageLSH}, an efficient index structure for processing image queries faster. \textit{bImageLSH} includes a bitmap-based performance optimization. For future work, we will look into finding theoretical bounds for our stopping conditions to make the theoretical foundation of \textit{bImageLSH} more robust. We will also investigate the effect of the underestimation and overestimation of the value of $\Gamma$ on different real datasets. Additionally, we will include more performance optimizations such as caching and improving the process of \textit{Virtual Rehashing}. 

\section{Conclusion}
Image data is often represented and stored as a collection of high-dimensional features.  Locality Sensitive Hashing is one of the most popular solutions for approximate processing in high-dimensional spaces, but it is not optimized to search for an image query (which is converted into a set of \textit{point queries}). In this paper, we presented an efficient index structure, \textit{bitmap-Image LSH (bImageLSH)} for efficient processing of images. Our novel design, which includes intuitive stopping conditions and a bitmap-based performance optimization, improves the performance of the image queries while keeping the accuracy high. Experimental results on a real dataset show the performance benefit of \textit{bImageLSH} (while keeping the accuracy high) when compared with the state-of-the-art LSH implementation.

\begingroup
\setstretch{0.83}

\endgroup

\end{document}